\documentstyle{article}
\input math_macros
\font\tenbf=cmbx10
\font\tenrm=cmr10
\font\tenit=cmti10
\font\elevenbf=cmbx10 scaled\magstep 1
\font\elevenrm=cmr10 scaled\magstep 1
\font\elevenit=cmti10 scaled\magstep 1

\renewenvironment{thebibliography}[1]
 { \elevenrm
   \begin{list}{\arabic{enumi}.}
    {\usecounter{enumi} \setlength{\parsep}{0pt}
     \setlength{\itemsep}{3pt} \settowidth{\labelwidth}{#1.}
     \sloppy
    }}{\end{list}}

\begin{document}
\begin{center}{{\tenbf MEASURING CHARM AND BOTTOM QUARK MASSES\\
(to appear in the DPF92 Proceedings)\\}
\vglue 1.0cm
{\tenrm THOMAS J. WEILER \\}
\baselineskip=13pt
{\tenit Department of Physics \& Astronomy, Vanderbilt University \\}
\baselineskip=12pt
{\tenit Nashville, TN 37235 USA\\}
\vglue 0.3cm
{\tenrm and\\}
\vglue 0.3cm
{\tenrm KRASOOS GHAFOORI--TABRIZI\\}
{\tenit Department of Physics, University of Shahid Beheshty\\
	Eween, Tehran 19834 Iran,\\
	and Center for Theoretical Physics and Mathematics\\
	Atomic Energy Organization of Iran, Tehran, Iran\\}

\vglue 0.8cm
{\tenrm ABSTRACT}}
\end{center}
\vglue 0.3cm
{\rightskip=3pc
 \leftskip=3pc
 \tenrm\baselineskip=12pt
 \noindent
The meaning and the
extraction of heavy quark masses are discussed.
A simple production model is presented which
incorporates the running of the
heavy quark mass into perturbative calculations.
The model offers the possibilities of (i) understanding
the differing charmed mass values extracted from different
experiments, (ii) determining the short--distance
mass relevent to quark mass matrix and mixing angle
studies, and (iii) determining the long--distance charm mass,
which determines the charm quark threshold and
sensitively affects the extraction of $\sin^2\theta_w $.
Threshold and forward angle
production offer the best possibilities to test the
model and extract meaningful charm/bottom masses.
\vglue 0.6cm}
{\elevenbf\noindent 1. Quark Masses and QCD}
\vglue 0.3cm
\baselineskip=14pt
\elevenrm
This year, the Particle Data Group (PDG) introduced into
the {\elevenit Review of Particle Properties} a ``Quark Table," in which
they list values for the quark masses$^1$.
The d--, u--, and s--quark mass values are
``current--quark masses" extracted from pion and kaon masses using chiral
symmetry.  The c-- and b--quark mass values are potential model masses
estimated from charmonium, bottomonium, D, and B masses; they are not
the running masses derivable from the QCD Lagrangian.
Moreover, the masses are poorly determined: $m_c$ is given a
range of 1.3 to 1.7 GeV, and $m_b$ is given a range of 4.7 to 5.3 GeV.
The PDG say that ``since
the subject of quark masses is controversial, the purpose of the table is
to provoke discussion."

Experiments on quark scattering and production can provide
the PDG with running QCD masses, with the running scale
provided by the subprocess invariants $\hat {s}$, $\hat{t}$, and $\hat{u}$.
Because of the running of QCD, one expects to find
(i) scattering quark masses smaller than the potential model
masses listed above, and (ii) extracted mass values that change with
scale and with reaction channel.  Point (i) expresses the fact tha QCD is
asymptotically free. At large scale one expects a measured mass to be
the bare current mass in the electroweak Lagrangian; this mass originates
from the Higgs mechanism and has nothing to
do with QCD.  Point (ii) reflects running, but also the fact that different
reaction channels have different intrinsic scales.  For example, in
Drell--Yan or $e^+e^-$ production, the quark lines are
external and the quarks are constituent--like; while in
heavy quark production via boson--boson fusion there is an internal quark
line and the associated quark is a short--distance, off--shell
(by $\hat{t}-m_Q$)
current quark.  Reactions with t-- or u--channel quark exchange will
yield lighter quark mass values than reactions without.

Unfortunately, the extraction of scattering masses is necessarily
model--de\-pen\-dent, for ``hadronization" or
``fragmentation" of the final state quarks is inherently
nonperturbative (hadrons and jets do not appear in the QCD Lagrangian), and
nonperturbative QCD must be modelled rather than calculated.
\vglue 0.5cm
{\elevenbf\noindent 2. Models for Heavy Hadron Production}
\vglue 0.3cm
One way to view the model dependence of the perturbative/nonperturbative
QCD interface is to ask,
at what stage in the calculation do nonperturbative effects enter?  In
conventional QCD phenomenology, a common mass parameter is used everywhere
in the Feynman diagrams and hadronization is added on in a
classical fashion.
The charmed mass value that emerges from fits to hadroproduction
data (where gluon--gluon fusion is dominant)
appears to us to be too large.  Fits with lowest order QCD give
$m_c=1.2\;GeV$, which is fine, but fits with loop--corrected QCD
give $m_c=1.5\;GeV$, which is as large as the mass determined from
potential models!
Furthermore, in a one--mass model there is no
possibility to understand the different mass values that seem to emerge
from different reaction channels.  And finally, there is no
possibility for running the mass into the nonperturbative region
where the running is greatest.

Thus, there is motivation to look at other models for the
perturbative/non\-per\-tur\-ba\-tive interface.  One simple approach
is to admit the heavier, dressed, constituent mass in the phase space
limits.  A more motivated two--mass model has recently been
introduced by us$^2$.  It models
running of the heavy quark mass at the Feynman diagram level:
the mass in
quark propagators is identified with the short--distance mass arising
from electroweak symmetry breakdown, and the mass in
the ``free" Dirac spinors and in the phase space limits
is identified with the long-distance/constituent mass.  Specifically,
quark propagators are$ ({\not\! p} - m_{SD})^{-1}$,
while quark spinors satisfy the
Dirac equations $ ({\not\! p}-m_{LD}) u(p,m_{LD}) = 0$ and
$({\not\! p}+m_{LD}) v(p,m_{LD}) = 0$.  SD and LD denote short and long
distance, respectively.  An immediate prediction is
that charm-masses extracted from reaction
channels dominated by graphs with (without)
internal charm-quark lines will have smaller (larger) values.

The LD constituent mass in the
Dirac spinor encompasses some of the nonperturbative
physics of color bleaching, fragmentation, and hadronization.
It may also be viewed as arising from a mass insertion on external quark
legs due to interactions with QCD vacuum condensates.
As such, it is a simple representation of highly complicated
physics.  The successes of
the nonrelativistic quark model in describing static
hadron properties argue that constituent quarks do
behave like Dirac particles, a result supported by current algebra$^3$.

Assigning different masses to internal and external lines creates
nonconserved currents, which break gauge invariance.  This becomes an issue
in higher order calculations where internal gauge boson lines are present.
The breaking of gauge invariance can be avoided
by retaining
$({\not\! p}-m_{SD}) u(p,m_{LD}) = 0$ and $({\not\! p}+m_{SD}) v(p,m_{LD}) =
0$.
Then the LD mass shows up only in the relations
$u(p,m_{LD})\overline{u}(p,m_{LD}) = {\not\! p}+m_{LD}$ and
$v(p,m_{LD})\overline{v}(p,m_{LD}) = {\not\! p}-m_{LD}$, and in the phase
space limits.  Alternatively, one may note that when
nonperturbative effects turn on, the physics that results
looks nothing like any known
extrapolation from the QCD Lagrangian.  Hence it may make sense to
allow nonperturbative effects to break gauge invariance in any perturbative
calculation, with the faith that an all orders calculation will produce the
exact gauge invariant physics.  (Just this philosophy is adopted in some
versions of light--cone QCD.)  This point of view motivates calculating in
physical gauges, where unitarity is manifest.  Further discussion on this
issue is contained in ref.2.
\vglue 0.5cm
{\elevenbf\noindent 3. Experimental Comparison of Heavy Hadron QCD Models}
\vglue 0.3cm
In ref.2 it is shown that $m_{SD}$ determines
the peak magnitude and the asymptotic magnitude of the subprocess
$gg\rightarrow  c{\bar c}$ cross section,
while $m_{LD}$ determines the threshold energy.  Thus, in principle both masses
are measureable.  The running-mass model gets both the
threshold and the rate correct with
$ m_{SD} \sim 1.2 $ GeV and $ m_{LD} \stackrel{>}{\sim} 1.5 $ GeV.
To quantitatively distinguish between conventional perturbative QCD and this
model, it may be necessary to compare across reaction
channels; we have mentioned that this
model predicts a lighter charm mass only
for those reactions having a t- and/or u-channel
charmed line.  It may also
be possible to distinguish between the
two models by examing a single reaction cross
section near threshold where the greatest differences in shapes occurs$^2$,
or near the forward scattering
angle, where $\hat t$ most closely approaches the SD
charmed-mass pole.  It may be possible to experimentally determine $\hat s$
and $\hat t$ (or $\theta_{cm}$) on an event by event basis,
through final state measurements. Eventually, photon-photon
charm-production data will become available; in this
reaction, $\hat s$ and $\hat t$ are measureable, and
there is no dependence on an initial state gluon
distribution.  The forward scattering peak is quite sensitive to
the charmed mass$^2$.

We encourage charm and bottom production
experimenters to analyze data with the model discussed here.
If it turns out that this model and the conventional QCD model
both fit the data and yield differing charm masses,
then available experiments are insufficient to quantitatively
determine \underline{the} short--distance charm mass.
But if a detailed study should show a preference
for one model over the other, then Nature will have spoken, and we will
have listened.
\vglue 0.5cm
{\elevenbf\noindent 4. References \hfil}
\vglue 0.3cm

\end{document}